\documentstyle[12pt,epsf,epsfig,amsmath]{article}
\newfont{\thiplo}{msbm10 scaled\magstep 2}
\newfont{\gothic}{eufb10 scaled\magstep 2}
\newfont{\unc}{eurb10} 
\newskip\humongous \humongous=0pt plus 1000pt minus 1000pt
\def\caja{\mathsurround=0pt}
\def\eqalign#1{\,\vcenter{\openup1\jot \caja
        \ialign{\strut \hfil$\displaystyle{##}$&$
        \displaystyle{{}##}$\hfil\crcr#1\crcr}}\,}
\newif\ifdtup

\def\eqright #1\cr{\noalign{\hfill$\displaystyle{{}#1}$}}
\def\eqleft #1\cr{\noalign{\noindent$\displaystyle{{}#1}$\hfill}}

\def\oldreffmt#1{\rlap{[#1]} \hbox to 2\parindent{}}

\def\figfmt#1{\rlap{Figure {#1}} \hbox to 1in{}}

\def\sectioneq{\def\theequation{\thesection.\arabic{equation}}{\let
\holdsection=\section\def\section{\setcounter{equation}{0}\holdsection}}}%

\newcounter{holdequation}

\def\auto{\eqno(\refstepcounter{equation}\theequation)}
\def\begineq #1\endeq{$$ \refstepcounter{equation}\eqalign{#1}\eqno
	(\theequation) $$}
\def\contlimit{\,{\hbox{$\longrightarrow$}\kern-1.8em\lower1ex
\hbox{${\scriptstyle (a\rightarrow0)}$}}\,}
\def\centeron#1#2{{\setbox0=\hbox{#1}\setbox1=\hbox{#2}\ifdim
\wd1>\wd0\kern.5\wd1\kern-.5\wd0\fi
\copy0\kern-.5\wd0\kern-.5\wd1\copy1\ifdim\wd0>\wd1
\kern.5\wd0\kern-.5\wd1\fi}}
\def\centerover#1#2{\centeron{#1}{\setbox0=\hbox{#1}\setbox
1=\hbox{#2}\raise\ht0\hbox{\raise\dp1\hbox{\copy1}}}}
\def\centerunder#1#2{\centeron{#1}{\setbox0=\hbox{#1}\setbox
1=\hbox{#2}\lower\dp0\hbox{\lower\ht1\hbox{\copy1}}}}
\def\lsim{\;\centeron{\raise.35ex\hbox{$<$}}{\lower.65ex\hbox
{$\sim$}}\;}
\def\gsim{\;\centeron{\raise.35ex\hbox{$>$}}{\lower.65ex\hbox
{$\sim$}}\;}

\def\super#1{\ifmmode \hbox{\textsuper{#1}}\else\textsuper{#1}\fi}
\def\textsuper#1{\newcount\holdspacefactor\holdspacefactor=\spacefactor
$^{#1}$\spacefactor=\holdspacefactor}

\def\getcite#1,{\advance\citenumber by1
\def\getcitearg{#1}\def\lastarg{@}
\ifnum\citenumber=1
\ref{#1}\let\next=\getcite\else\ifx\getcitearg\lastarg\let\next=\relax
\else ,\ref{#1}\let\next=\getcite\fi\fi\next}

\def\pom{{\rm P\kern -0.53em\llap I\,}}
\def\spom{{\rm P\kern -0.36em\llap \small I\,}}
\def\sspom{{\rm P\kern -0.33em\llap \footnotesize I\,}}

\relax
\def\contlimit{\,{\hbox{$\longrightarrow$}\kern-1.8em\lower1ex
\hbox{${\scriptstyle (a\rightarrow0)}$}}\,}
\def\upon #1/#2 {{\textstyle{#1\over #2}}}
\relax
\renewcommand{\thefootnote}{\fnsymbol{footnote}}

\sectioneq

\def\til#1{\centeron{\hbox{$#1$}}{\lower 2ex\hbox{$\char'176$}}}
\def\tild#1{\centeron{\hbox{$\,#1$}}{\lower 2.5ex\hbox{$\char'176$}}}
\def\sumtil{\centeron{\hbox{$\displaystyle\sum$}}{\lower
-1.5ex\hbox{$\widetilde{\phantom{xx}}$}}}

\begin{document} 

\begin{titlepage} 

\rightline{\vbox{\halign{&#\hfil\cr
&ANL-HEP-PR-04-47\cr
&\today\cr}}} 
\vspace{0.25in} 

\begin{center} 
  
{\large\bf Sextet Quark Physics at the Tevatron ? }\footnote{Work 
supported by the U.S.
Department of Energy under Contract
W-31-109-ENG-38} 

\medskip

Alan. R. White\footnote{arw@hep.anl.gov }

\vskip 0.6cm

\centerline{Argonne National Laboratory}
\centerline{9700 South Cass, Il 60439, USA.}
\vspace{0.5cm}

\end{center}

\begin{abstract} 

A sextet quark sector of QCD, together with the sextet higgs mechanism,
would produce a major change in the strong interaction above the 
electroweak scale. This change may already be evident in Cosmic Ray physics and, if so,
dramatic effects are to be expected at the LHC. In this paper 
we discuss whether evidence for the sextet sector could be seen at the Tevatron.

A major consequence of the connection between QCD and electroweak symmetry
breaking is the strong coupling of the pomeron to the electroweak sector.
At the Tevatron, the energy is too low to produce vector boson pairs
directly via double pomeron exchange, but a number of small cross-section 
effects could be seen in diffractive, and diffractive related, processes involving 
$W^{\pm}$ and $Z^0$ vector bosons. Probably, the most important feature that 
could be decisively established is that the production cross-section for 
$W^+W^-$ and $Z^0Z^0$ pairs has an anomalous component with event 
characteristics different from the Standard Model. This would be the first 
indication of what should become a dominant, very large, cross-section
at the LHC.

If the sextet quark dynamical mass scale is well above the top quark mass, then 
the production properties of $W$'s and $Z$'s could 
be the only new physics visible at the Tevatron scale. If this scale is  
lower the situation could be more subtle.   
The observed $t\bar{t}$ events could  
originate from the $\eta_6$ - the ``sextet higgs'', even though they 
can be understood perturbatively. The interpretation of the top quark mass 
would, however, be different and non-perturbative decay modes should also be seen. 
A jet excess at large $E_T$  would provide 
supporting evidence for this picture, since
$\alpha_s$ evolution should stop at $E_T \sim$ ``$m_{top}$''.

\end{abstract}

\renewcommand{\thefootnote}{\arabic{footnote}} \end{titlepage}

At first sight, the ``sextet higgs mechanism'', in which
the $W^{\pm}$ and $Z^0$ aquire 
their mass via the QCD chiral symmetry breaking of a color sextet quark sector,
can be regarded\cite{bww} 
as a very particular form of the general technicolor 
higgs mechanism\cite{ta}, with QCD as the 
technicolor gauge group. Indeed, 
this is the framework in which it was originally proposed\cite{wm}.
If we consider the Standard Model without the usual higgs sector and add
a doublet of sextet quarks ($U$ \& $D$), the 
breaking of the sextet chiral symmetry gives a triplet 
of ``sextet pions'' ($\Pi^{\pm}$, $\Pi^0$) and
a ``sextet higgs'' particle - the $\eta_6$. The $W^{\pm}$ and 
$Z^0$ then aquire masses by ``eating'' the $\Pi$'s.

In fact, as we emphasized in \cite{arw03} and will emphasize further 
in a new paper that is in preparation\cite{arw04},
the sextet mechanism is radically different
from the technicolor mechanism (and, in effect, all other proposed higgs
mechanisms). Firstly, because the electroweak
scale is, economically and very beautifully, a second (higher color) QCD scale, 
no new interaction is needed beyond the familiar 
SU(3)xSU(2)xU(1) gauge interactions of the Standard Model.
(A short-distance
SU(2)xU(1) anomaly can be avoided by adding heavy leptons,
although an underlying grand unified theory could give a more 
elaborate high mass cancelation\cite{awkk}.)
Secondly, electroweak symmetry breaking is intricately connected
with QCD dynamics and, most importantly, produces
a major change in the strong interaction above the electroweak scale. 
As we will briefly discuss, we believe this change is already evident in very 
high-energy Cosmic Ray physics. If this is the case, then large cross-section 
physics must necessarily be involved and it  
will be inescapably apparent at the LHC. At the 
Tevatron the situation could be more ambiguous.
In this paper we will describe, in a general manner, 
some sextet quark physics that might be seen at the Tevatron.\footnote{We  
also discussed expected sextet sector effects, at accelerators\cite{arw97}, 
and in Cosmic Ray physics\cite{arw94},
previously. Since then, however, an improved
understanding of $QCD_S$ has led us to different expectations,
particularly concerning the irrelevance of instanton interactions.}  

We refer to QCD with six triplet flavors and two sextet flavors
as $QCD_S$. (The suffix can be thought of 
as denoting either ``sextet'' or ``special'', or even ``saturated'', since the 
asymptotic freedom constraint on the quark content of QCD is saturated.)
$QCD_S$ has several special features and,
for a long time\cite{arw02}, we have argued that it 
has the particular attraction that it gives Critical Pomeron asymptotic high energy
behavior\cite{cri} (uniquely satisfying unitarity in all aspects).
We have also argued that the pomeron and infinite momentum hadron states
emerging from our work correspond to a special (S-Matrix) solution
of QCD that is very close to perturbation theory
and appears only in $QCD_S$. 

Because of the particular quark content of $QCD_S$,
the high-energy behavior can be constructed from the reggeon diagrams of 
the color superconducting version of the theory (in which the gauge symmetry
is broken from SU(3) to SU(2) ). The key feature of the 
superconducting theory is that a
``wee gluon anomaly condensate'' is produced by the interplay 
between the chiral anomaly properties of Goldstone bosons and 
reggeon vertex anomalies\cite{arw03,arw02}. 
In $QCD_S$ the wee gluon condensate becomes 
a ``dynamical wee gluon component'' of infinite momentum physical states that,
in effect, is responsible for the non-perturbative properties 
of confinement and chiral symmetry breaking\footnote{We will discuss in \cite{arw04}
how construction of $QCD_S$ via the superconducting theory resolves the 
infinite momentum quantization ambiguity of the contribution  
of ``unphysical'' longitudinal wee gluons, an ambiguity that is well-known to be 
related to the finite momentum choice of vacuum.}.

The origin of the wee gluon component, in the superconducting theory, implies
that it carries both (global) color and spin. Also, since all ``hadrons'' originate
as Goldstone boson states in the superconducting theory, they necessarily have a 
short-distance (large momentum) 
component that is gauge invariant (via reggeization) but 
retains global color and spin properties that are canceled by the wee gluon
component. A ``pion'' 
is, in first approximation, a color octet quark/antiquark
pair (either color triplet or color sextet)
in a vector-like spin state combined with a wee gluon component. Similarly, 
the pomeron (in first approximation) is a color octet (reggeized) vector gluon 
combined with a wee gluon component. Because of the anomaly origin of the wee gluon
component it can not be generated (radiatively) by the
short distance component. As a result, there is a well-defined separation
between the short-distance and wee gluon components
which implies that a 
parton model is valid. However, the special features
of the short-distance component are different to those usually anticipated in 
the QCD parton model. Most importantly, there is no short-distance contribution
that carries directly all the quantum numbers of the hadron. 
This is crucial for the status of the $\eta_6$ in $QCD_S$, as we will discuss
later.

Because the (high-energy) solution of $QCD_S$ is so close to perturbation theory,
dynamical triplet and sextet quark momentum scales 
will be related (approximately) by the ``Casimir Scaling'' rule
that is, roughly, satisfied by Feynman diagrams. For example, if $F_{\pi}$ and
$F_{\Pi}$ are, respectively, triplet and sextet chiral scales, we expect
$$
C_6~\alpha_s (F_{\Pi}^2)~\sim ~C_3 ~\alpha_s(F_{\pi}^2)
~~~~~~~~~~~C_6/C_3 ~\approx~ 3
\auto\label{CSC}
$$
where $C_3$ and $C_6$ are triplet and sextet Casimirs.
If $\alpha_s$ evolves sufficiently slowly (e.g. $
\alpha_s (F_{\pi}^2)
\sim 0.4~$) then $F_{\Pi}$, which gives the mass scale for $W^{\pm}$ and $Z^0$
vector bosons, can indeed be the electroweak scale. 

As we discussed in \cite{arw03}, and will discuss 
at length in \cite{arw04}, $F_{\Pi}$ also provides the scale for the coupling
of the wee gluon component of the pomeron to sextet pions. 
Combining this with the formation of Goldstone boson
``pions'' via triangle diagram anomaly poles,   
we have developed\cite{arw03,arw04} 
a semi-perturbative method for estimating the magnitude
of hard diffractive production of $W$'s and $Z$'s 
when the sextet higgs mechanism is operative. 
The essential feature is that the diffractive interactions
shown in Fig.~1 can be shown to be large, via anomaly pole production, 
when $k_{\perp}~ \centerunder{\raisebox{0.05mm}{$>$}}{$\sim$}~ M_W$. 
\begin{center}
\epsfxsize=5in
\epsffile{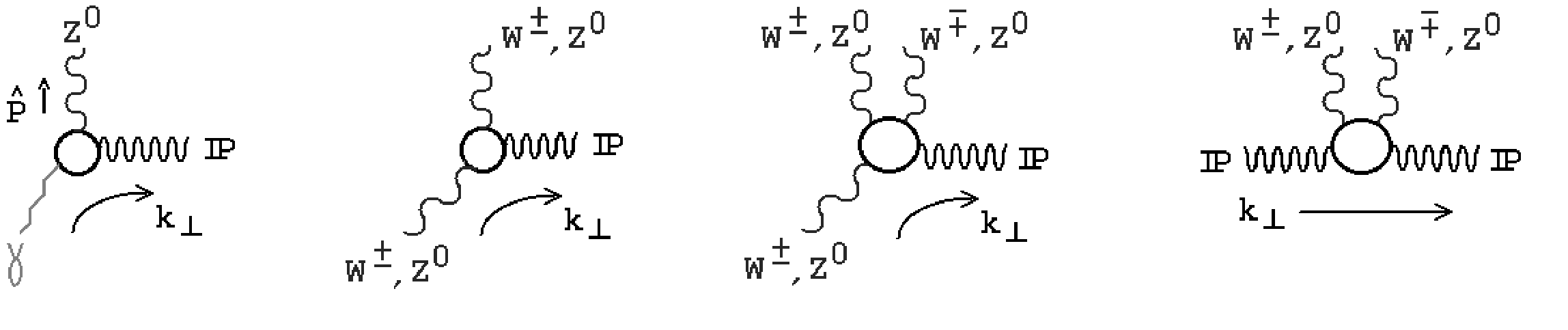}

Fig.~1 Diffractive Interactions Produced by Sextet Quark Loop Anomalies 
\end{center}

The first two interactions, in Fig.~1, contain a hard vertex which combines 
with a sextet quark loop anomaly pole, involving the wee gluons in the pomeron, to
produce the sextet pion that becomes a final state vector boson. 
In the second two interactions, two sextet pions are produced
and so there are two sextet quark loop anomalies within each interaction.
The hard diffractive estimates can also be combined with pomeron regge theory
to obtain predictions for soft diffraction.
The anomaly pole method allows us to estimate 
the interactions at large $k_{\perp}$ and, when continued to
smaller momentum transfers, the rapid increase of the pomeron coupling
to a scattering hadron state strongly enhances the cross-section.

In \cite{arw03} we briefly discussed how, at HERA, 
events\cite{HE} at large $x$ and $Q^2$ might be produced (essentially) 
by the first diffractive interaction in Fig.~1, in which a photon is excited
to a $Z^0$. At the Tevatron, the second interaction shown in Fig.~1
would allow a perturbatively produced
$W^{\pm}$ or $Z^0$ to scatter via pomeron exchange. This scattering could  
explain the push towards larger rapidities, that may have been 
observed\cite{d0}, when 
a $W^{\pm}$ or $Z^0$ is produced in association with a large $E_T$ jet. 
The third interaction shown in Fig.~1 should produce a 
diffractive cross-section for the production of $W^{\pm}$ and $Z^0$ pairs
that is large compared to the Standard Model diffractive cross-section.
This process
might be the most direct way to detect the presence of sextet quark physics 
at the Tevatron. Unfortunately, the size of the cross-section
is limited by requiring the initial, perturbative, production
of a $W$ or $Z$. 

The last interaction shown in Fig.~1 is a double pomeron interaction 
that does not require any initial vector boson production. As a result  
we argued in \cite{arw03} that, because of the enhancement 
by pomeron couplings at small momentum transfer, the LHC cross-section for
double pomeron production of $W^{\pm}$ and $Z^0$ pairs should be 
very large. This would be clear, direct, evidence for the sextet higgs mechanism 
that could be produced very soon after the LHC turns on. 

In fact, if the double pomeron interaction of Fig.~1 is large, then 
``cut pomeron'' amplitudes involving $W^{\pm}$ or $Z^0$ 
pairs coupling as sextet pions 
should also be large. In particular, the amplitude shown in Fig.~2, 
which describes the central region inclusive production of 
a $W^{\pm}$ pair, should be large, as should be the corresponding $Z^0$ pair
amplitude. As a result,
$W^{\pm}$ and $Z^0$ pairs should be multiply produced (more and more abundantly as 
the energy increases) across most of the rapidity axis, in close analogy
with pion production at much lower energies. 
\begin{center}
\epsfxsize=2.7in
\epsffile{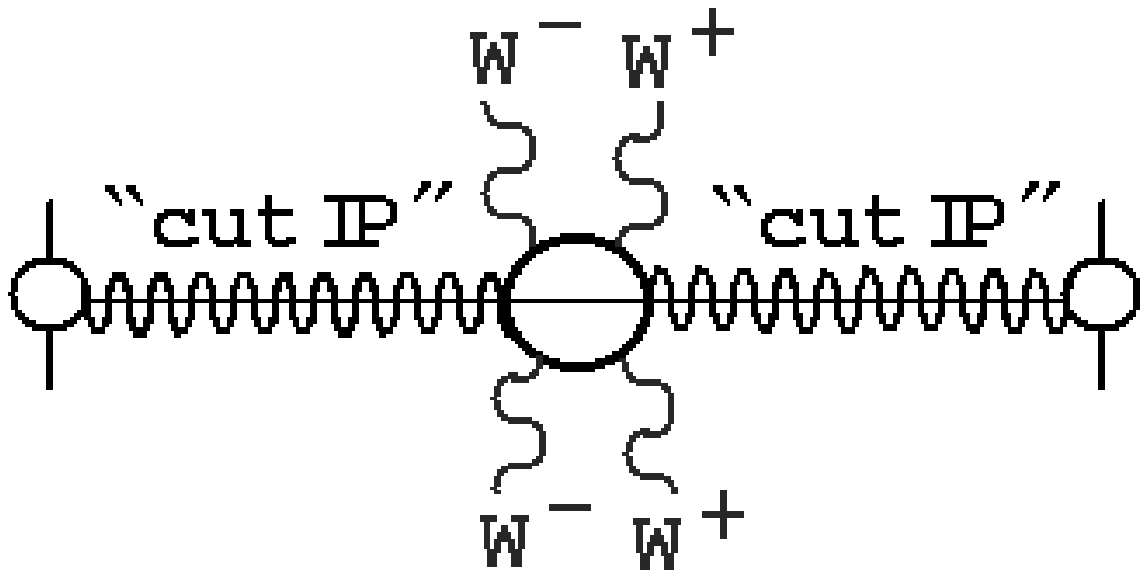}

Fig.~2 A Contribution to the $W^+W^-$ Inclusive Cross-Section. 
\end{center}
We believe this is a major change in the strong interaction which 
is able to explain the observed ``knee'' in the Cosmic Ray spectrum
(between Tevatron and LHC energies) that, while becoming more and more firmly
established, has baffled Cosmic Ray physicists for more than forty years\cite{swo}.

In the initial cosmic ray collision, because of the interaction change, 
the average transverse momentum of the produced particles
will rise dramatically with energy and an increasingly 
larger fraction will be undetected at ground level. Also a significant
fraction of the produced energy will go into neutrinos that are not detected.
As a result of these effects, 
the energy of the incoming Cosmic Ray will be seriously underestimated. The
result will be an, apparent, knee 
in the deduced incoming energy spectrum at
(roughly) the effective threshold for significant production of 
$W^{\pm}$ and $Z^0$ pairs. To produce as large an effect as is seen,
a significant part of the cross-section must be involved. In effect, 
at high enough energy, $W$ and $Z$ production must be competitive with pion production
in the strong interaction, which is not unreasonable if 
two comparable quark sectors of QCD are involved.  
Note that there is specific evidence\cite{cores} from 
the Cosmic Ray experiments of very large transverse momenta,
in ``dijet'' events, above the knee. The cross-sections involved 
are orders of magnitude larger than anticipated in conventional QCD.
Most probably, the dijets are $W$ or $Z$ pairs.

Apparently\cite{UA1}, there is already an anomalously large $W$ pair 
cross-section at the energy of the $S\bar{p}pS$ collider and since we expect
this cross-section to be really large at the LHC, it seems that an
``anomalous'' (although still relatively small) cross-section 
should surely be observed at the Tevatron. A complication is that detection of
events in which one of the pair decays hadronically is much more difficult 
at the Tevatron than it was at the $S\bar{p}pS$ because of the large background 
from the QCD production of $W$ (or $Z$) plus two jets.
Double pomeron production which, as we described above,
we expect to be a very clean signal
at the LHC, is (strictly) inaccessible kinematically at the Tevatron. 
However, the single diffractive interactions  
\newpage
\noindent 
discussed above, and other related events with unexpectedly 
low\footnote{Low multiplicity events can anticipate higher energy 
rapidity gap cross-sections.} 
(and high\footnote{A connection between 
diffractive cross-sections and events with twice the average multiplicity density
(in rapidity) is required by the AGK cutting rules. In addition, 
the Wilson lines attached to sextet quarks 
should generate higher associated multiplicities than triplet quark lines.}) 
associated multiplicity, should still give an anomalous cross-section that develops
into the anticipated very large cross-section at higher energies.

It is also, of course, very interesting to consider whether we can find evidence of
sextet quarks in jet physics, possibly
in events where no electroweak bosons are present. 
The contribution of one sextet quark to the $\beta$-function 
of QCD is the same as five triplet quarks. As a consequence, the inclusion of
a sextet quark doublet halts the evolution of $\alpha_s$ almost entirely. 
(Asymptotic freedom still holds, but it would be manifest only as a very 
slow decrease of $\alpha_s$ at energies well above the sextet scale.) Therefore,
if the jet inclusive cross-section is calculated using standard perturbative
QCD, $\alpha_s$ should not evolve beyond the $E_T$ scale at which the sextet quarks
enter the theory. If we consider just the sextet pions, i.e. the $W^{\pm}$ and $Z^0$,
then  $E_T \sim 2 M_W$ would appear to be  
the scale at which sextet quarks enter the (sextet) flavor 
neutral part of the theory. This is, indeed, roughly the scale 
above which the non-evolution of $\alpha_s$ can be viewed, experimentally, as 
responsible\cite{CDF} for a large $E_T$ jet excess. However, $M_W$ is the chiral 
scale of the sextet sector and, a priori, we would 
expect that the non-chiral dynamical 
mass scale would determine the scale at which
the evolution of $\alpha_s$ is affected. As we now discuss, we expect
this scale to be associated with the mass of the $\eta_6$.

Usually, it is assumed that the only explicit, non-perturbative, axial
chiral symmetry breaking in QCD is that 
due to topological (instanton) contributions to the anomaly current.
If this were the case in $QCD_S$, there would be a
U(1) symmetry associated with the $\eta_6$ that 
would be unbroken. As a consequence, in addition to being the
analog of the usual higgs scalar, the $\eta_6$ would also be\cite{cllr} 
a light axion of the kind that is ruled out experimentally. 

At this point, the existence of the 
wee gluon component of the $\eta_6$ is essential.
As we briefly noted in \cite{arw03}, and will discuss in 
more detail in \cite{arw04}, in $QCD_S$ the 
anomaly vertices that create the wee gluon 
component of a hadron break both the sextet and triplet U(1) symmetries.
As a consequence, there is no light axion in the spectrum.
Because of the presence of the wee gluons,
the short-distance component of the $\eta_6$ 
does not carry all of it's quantum numbers. Rather this component
is a flavor singlet
sextet quark/antiquark pair that carries 
octet color and so can couple directly to an, unphysical, component of 
the gluon. As a result,
a multigluon regge exchange (initially a daughter of the pomeron)  
mixes with the $\eta_6$. We anticipate that this mixing 
generates an electroweak scale dynamical mass for the $\eta_6$. 
(As we discuss below, the multigluon exchange also mixes
with the corresponding flavor singlet composed of color triplet quarks).

Our construction of high-energy $QCD_S$ 
is crucially dependent on effects of the chiral anomaly when the quarks are 
massless and so we necessarily construct  
the massless version of the theory first. In fact, in \cite{arw03}
and \cite{arw04} we  
discuss how vector boson masses are generated by the sextet higgs mechanism, 
but do not consider whether the generation of effective (current) triplet
quark masses could involve 
the same mechanism. To be physically applicable, therefore,
effects of triplet quark current masses, including that of the top, 
must be added to the S-Matrix of $QCD_S$. 
For the sextet higgs mechanism to be operative, sextet quarks can not have 
a current quark mass. How 
dynamically generated sextet mass scales 
compare with the top quark mass is, therefore, crucial.

It is possible that the $\eta_6$ mass could be too 
large for it to be seen at the Tevatron (say 1 TeV). In this case we could,
perhaps, assume that all effects of
the sextet sector, apart from those in which sextet pions are involved, can be 
integrated out at the top quark mass scale. This would imply
that standard perturbative
QCD could be applied to top quark production and the only evidence of the 
sextet quark sector would be in the vector boson
production cross-sections that we have already discussed. For consistency
$\alpha_s$ should, presumably, evolve as usual up to and beyond the top mass.
Therefore, a jet 
excess would have to be entirely due to the multiple production of $W$'s and $Z$'s
that would be outside of the standard perturbative calculation.

Even if the mass scale is very high, the existence of a non-perturbative 
QCD sector above the mass of the 
top quark makes it worrisome that the concept of a perturbative, electroweak
scale, current quark 
mass can be well-defined enough to be directly measured. (There would surely be
a large dynamical mass generated above, if not at, the electroweak scale.)
This worry becomes ever 
stronger as we decrease, as an assumption, the mass of the $\eta_6$. To discuss
what we might expect, it is instructive to consider the effect of smoothly adding
a top quark mass to massless $QCD_S$. 

In the massless theory the $\eta_6$ will
mix, as we noted above, with both a multigluon state and
the flavor neutral triplet quark meson (composed of all six triplet
quarks) that we will refer to as the $\eta_3$. Initially, because of their 
different dynamical mass scales, one state will remain primarily triplet
quark, which we can continue to identify as the $\eta_3$, while the other will
remain primarily a sextet state, and can be identified as the $\eta_6$.
Consider, first, the effect on the $\eta_3$ of increasing the top mass, ignoring 
the mixing with the $\eta_6$. As 
the top mass is increased the mass of the $\eta_3$ will 
increase, it's triplet quark content will become primarily $t\bar{t}$, and it 
will also become increasingly unstable - that is it's width will increase.
When the current quark top mass is such that the mass of the $\eta_3$ is well past
the threshold for $W^+W^-~b\bar{b}$ production
so that, effectively, the top quark has become significantly unstable, we 
anticipate that the $\eta_3$ will have such a large width that it will be 
physically unobservable. At this stage we would also anticipate that
the sextet sector is already contributing, dynamically, to the 
effective top quark mass.

If we now consider the fate of the $\eta_6$, we will find that the mixing with 
the $\eta_3$ increases as the top mass is increased. However, the essential
sextet quark composition of the $\eta_6$ will make it's mass higher.
It's width, which will come amost entirely from the mixing with the
triplet sector, will be narrower (although still large).
As a consequence the $\eta_6$ could be observable at a 
relatively low electroweak scale mass and have $t\bar{t}$
as a primary decay mode, in addition to other
triplet quark decay modes and non-perturbative $W$ and $Z$ decay modes 
that we discuss next. At the parton model level,
the $\eta_6$ would be produced primarily via gluon production, just as is the top.
It is natural, therefore, to raise
the possibility that the observation
of a $t\bar{t}$ ``threshold'' at the Tevatron might 
actually be the observation of the $\eta_6$. Since many experimental features would
be similar to the perturbative picture, a key signal of this 
could be the observation of, one or more, non-perturbative decay modes.

To discuss non-perturbative decay modes of the $\eta_6$, the best we can do
is to exploit the parallel between the \{$\Pi^{\pm},\Pi^0,\eta_6$\} sextet
states, corresponding to \{$W^{\pm},Z^0,\eta_6$\},
and the familiar \{$\pi^{\pm},\pi^0,\eta$\} triplet quark states.
Although the width would be large, we should, presumably, take the
mass of the $\eta_6$ to be $\sim~$``$~2 m_{top}~$'' $\sim$ 350 GeV. 
In this case, the relative couplings and masses of
the vector mesons, and the photon, imply that the 
primary non-perturbative decay mode should be (in parallel with 
$\eta~\to~ \pi^+~\pi^-~\pi^0$) 
$$
\eta_6~~\to~~ W^+~W^-~Z^0 
\auto\label{dk1}
$$
which, when $Z^0 \to b\bar{b}$, would give the same final state as $t\bar{t}$. 
The next most significant mode 
$$
\eta_6~~\to~~ Z^0~Z^0~Z^0 
\auto\label{dk2}
$$
(in parallel with $\eta~\to~ \pi^0~\pi^0~\pi^0$) 
should have a smaller branching ratio, because of the larger $Z^0$ mass. 
In addition, (\ref{dk2}) would  
be indistinguishable from (\ref{dk1}) when the $Z^0$'s decay hadronically, as they
do most of the time. Because the $\eta_6$ mass is so large, 
decay modes involving an electromagnetic coupling, such as
$$
\eta_6~~ \to~~ W^+~W^-~\gamma~, ~~~Z^0 ~Z^0~\gamma ~, ~~~ Z^0 ~\gamma~\gamma~, 
~~~ \gamma~\gamma
\auto\label{dk3}
$$
would be expected to have smaller branching ratios but should 
be present at some level. 

If the top quark events are produced by the $\eta_6$ then ``$~m_{top}~$''
would be the sextet dynamical mass scale above which $\alpha_s$ would not
evolve. In this case there should surely be a jet excess at 
the Tevatron which, at least in part, can be interpreted\cite{CDF} as non-evolution 
of $\alpha_s$ beyond $E_T \sim$``$~m_{top}~$''. Although there is little difference 
between $E_T \sim 2 M_W$, as discussed above, and $E_T \sim$``$~m_{top}~$'',
it is clear that if the top mass has the significance that we are now
discussing then the sextet sector has fully entered the theory at this scale.  
Note that the increasing entry of sextet sector states into the dynamics
should imply that the ``excess'' continues to grow as $E_T$ increases.
Indeed, we would expect that in the highest $E_T$ excess region there is an 
enrichment of jets with $M_{jet} \approx M_{W/Z}$.

Theoretically, and ``philosophically'', it would surely be attractive if
an electroweak scale mass, i.e. 350 GeV, is explained
as the (dynamical) mass of a sextet
quark/antiquark bound state, rather than as (twice the value of) a 
lagrangian parameter of the triplet
quark sector. Whether a well-determined top quark
``mass'' should still be, experimentally, identifiable is not clear. 
Theoretically, it would also be appealing if 
the logical paradox, that the mass of a 
colored (confined) state is a well-defined physical observable, 
could be avoided altogether. 

Since top quark physics at the Tevatron is very complex, with 
elaborate analyses needed to make a connection between theory and experiment, it 
is clear that it may not be an easy place to look for new physics
of the kind we have discussed. Whether or not 
$W^+W^-$ and $Z^0Z^0$ production conform to Standard Model expectations may be
a much more straightforward issue to determine. If, however, significant 
evidence for 
sextet quark physics begins to accumulate then, obviously, 
all possible discovery directions should be pursued intensely. 

{\bf Acknowledgement}

I am indebted to Mike Albrow for stimulating the serious consideration of the role 
of the $\eta_6$ that led me to realize what it's true significance might be.

\end{document}